\documentclass[12pt,a4paper]{article}

\newcommand{\be}{\begin{equation}}
\newcommand{\ee}{\end{equation}}
\newcommand{\bea}{\begin{eqnarray}}
\newcommand{\eea}{\end{eqnarray}}

\begin{document}

\title{ Gauge Backgrounds from Commutators and the Role of Noncommutative Space}

\author{Ciprian Sorin Acatrinei \\
         \\
	Theory Division, National Institute for Nuclear Physics \\
	and Engineering, P.O. Box MG6, Bucharest, Romania \\
	and \\
        Department of Physics,  University of Crete \\
        P.O. Box 2208, Heraklion 71003, Greece \\
        {\it e-mail: acatrine@theory.nipne.ro}  }
\date{February 3, 2004}

\maketitle

\begin{abstract}

The gauge connections corresponding to electromagnetism, Yang-Mills theory and Einstein gravity
can be derived by assuming specific commutation relations between the phase-space variables 
of a first quantized theory. 
Extending the procedure to noncommuting coordinates leads to new types of dynamics, which are explored. 
In particular, the conditions for the coexistence of an electromagnetic background
and a noncommutative two-from are found, as well as a generalized mechanism of dimensional reduction.
The noncommutative deformation of a gravitational background is also constructed.
The present formulation suggests some simple experimental tests of noncommutativity.

\end{abstract}

\section{Introduction and summary}

Gauge invariance appeared as a useful symmetry of the Maxwell equations. 
It gradually gained recognition, especially after the advent of quantum mechanics. 
In 1929, Weyl \cite{weyl} proposed it as a basic postulate, 
which enforced the existence of electromagnetic interactions.
The subsequent generalization to non-Abelian \cite{ym} and gravitational
\cite{utiyama} interactions, together with the ideas of spontaneous symmetry breaking 
and confinement, confirmed the universality of the gauge principle.

The first part of this paper endeavours to present a slightly unconventional viewpoint on gauge fields.
It shows in a unified way that part of the content of a gauge theory 
- the connection and its corresponding field strenght - 
follow by postulating appropriate nonzero commutators between the 
momenta, or momenta and coordinates, of a first-quantized theory. 
The equations of motion of a point particle subject to such a gauge background also follow.
When the underlying space is commutative, the above commutators fall naturally into three classes, 
corresponding to electromagnetic, Yang-Mills and gravitational fields.
A brief description of a matrix model-like formulation for gauge fields also appears in Section 2.

Section 3 - the main part of the paper - extends the analysis to 
nontrivial commutators between the coordinates, 
and discusses the physical consequences of such situations.
The conditions under which an electromagnetic-like background and noncommutative
space can coexist, for instance, are quite restrictive. They lead to
simple signals of noncommutativity, which could be detected in a low-energy, cummulative 
and relatively simple experiment.
A generalized mechanism for dimensional reduction is also discussed.
A gravitational background on the other hand is not restricted in any way by noncommutativity,
although the equations of motion of the test particle become more complex. 
They are written down for small noncommutativity, and can be used to put bounds on the noncommutative scale.
A discussion of the significance of the noncommutativity 2-form closes Section 3.

\section{Standard backgrounds via commutators}

\subsection{Electromagnetism and Yang-Mills}

The simplest instance in which the commutation relations imply the existence
of a gauge connection and symmetry is electromagnetism.
To begin, consider a nonrelativistic $(d+1)$-dimensional quantum-mechanical theory. 
Two elements encode the dynamics: the Hamiltonian, $\hat{H}(\hat{q}_i,\hat{p}_i)$, 
and the commutation relations between the phase space variables $\hat{q}_i$ and $\hat{p}_i$.
We will use a free  Hamiltonian, $\hat{H}=\frac{1}{2m}\hat{p}^2$,
and will generate nontrivial dynamics through the following commutation relations: 
\be
[q_i,q_j]=0 \qquad [q_i,p_j]=i\delta_{ij} \qquad [p_i,p_j]=iF_{ij}(q_k). \label{cr1}
\ee
The $[p,p]$ commutator implies the presence of a magnetic field strength $F_{ij}$.
In the Schr\"odinger picture, this is seen as follows.
In the coordinate basis $|q_i>$ the relations (\ref{cr1}) are represented by
($\frac{\partial}{\partial q_i}\equiv\partial_{q_i}\equiv\partial_i$)
\be
\hat{q}_i|q_i>=q_i|q_i>, \qquad \hat{p}_i|q_i>=(-i\partial_{q_i}-A_i(\vec{q}))|q_i>,
\ee
provided
\be
\partial_{q_i}A_j-\partial_{q_j}A_i=F_{ij}(q_k). \label{curl}
\ee
The gauge connection appeared as a consequence of the nonvanishing of the $[p,p]$ commutator.
Eq. (\ref{curl}) is not an assumption; it is enforced by the Jacobi identity for $p_i,p_j,p_k$, 
\be
\partial_i F_{jk}+\partial_k F_{ij}+\partial_j F_{ki}=0, \label{Jacobi_F}
\ee
which is equivalent to the sourceless Maxwell equations.
The Jacobi identity also implies that
\be
[[p_j,p_k],q_i]=-i\frac{\partial F_{jk}}{\partial p_{i}}=0,
\ee
precluding any dependence of $F$ on $p$. Thus $F(q)$ is a necessary condition, not an assumption.
Due to (\ref{Jacobi_F}), the present formalism does not accomodate magnetic monopoles.
The Schr\"odinger equation includes a gauge field,
\be
\hat{H}\Psi(q_i)=-\frac{1}{2}(-i\partial_{i}-A_i)^2\Psi(q_i)=E\Psi(q_i), \label{S_em}
\ee
and gauge invariance is built in automatically. 
%
In operatorial formulation, the Heisenberg equations of motion read 
\be
\dot{q}_i=-i[q_i,H]=p_i \qquad \dot{p}_i=-i[p_i,H]=\frac{1}{2}(F_{ij}p_j+p_jF_{ij}), \label{HemF}
\ee
and lead to 
\be
\ddot{q}_i=\frac{1}{2}(F_{ij}\dot{q}_j+\dot{q}_jF_{ij}).
\ee 
This is the operatorial Lorentz force law for a particle in a magnetic field.
The usual $\ddot{q}_i=F_{ij}\dot{q}_j$ law is obtained by prescribing 
an ordering for the noncommuting operators $q_i$ and $p_i$. 
If one works at the classical level, with Poisson brackets instead of commutators, 
one obtains directly $\ddot{q}_i=F_{ij}\dot{q}_j$.

The above considerations extend to 
relativistic quantum mechanics. Consider the operators $p_{\mu}$ and $q_{\mu}$, $\mu=0,1,2,\dots,d$; 
$q_0=t$ is treated formally as an operator, canonically conjugated to the Hamiltonian $p_0=H$. 
If
\be
[q_{\mu},q_{\nu}]=0 \qquad [q_{\mu},p_{\nu}]=i\eta_{\mu\nu} 
\qquad [p_{\mu},p_{\nu}]=-iF_{\mu\nu}(q),
\ee
(with metric $\eta=Diag[-++\cdots+]$) one has, 
in the $|q_0, \vec{q}>$ basis,
\be
\hat{q}_{\mu}|q_{\sigma}>=q_{\mu}|q_{\sigma}>, \qquad 
\hat{p}_{\mu}|q_{\sigma}>=[i\partial_{\mu}+A_{\mu}(q_{\nu})]|q_{\sigma}>,
\ee
\be
\partial_{q_{\mu}}A_{\nu}-\partial_{q_{\nu}}A_{\mu}=F_{\mu\nu}(q_{\rho}),
\ee
the later enforced again by the Jacobi identity.
The Schr\"odinger equation follows from $2mH\Psi=\vec{p}^2\Psi$,
and the Klein-Gordon equation in an electromagnetic background from $H^2\Psi=(\vec{p}^2+m^2) \Psi$.
A more elegant way is to introduce the proper time $\tau$, on which $p_{\mu}, q_{\nu}$ depend.
By studying the proper-time evolution of $p_{\mu},q_{\mu}$ under the fictitious Hamiltonian
$2\bar{H}_{\tau}=p_{\mu}^2$, one obtains the particle equation of motion 
$\frac{d^2{q}_{\mu}}{d\tau^2}=\frac{1}{2}(F_{\mu\nu}q_{\nu}+q_{\nu}F_{\mu\nu})$.

The commutation relations fix the sourceless 
part of the Maxwell equations (the Bianchi identity, or existence of a connection), 
together with the equation of motion of a particle situated in an electromagnetic field (Lorentz law).
The Maxwell equations with sources require extra assumptions, 
but can be inferred quite naturally: the Schr\"odinger equation (\ref{S_em}) 
displays a divergenceless current $j_{\mu}$, which shares this property with $\partial_{\nu}F_{\mu\nu}$. 
This suggests that, up to a multiplicative factor,
\be
\partial_{\mu} F_{\mu\nu}=j_{\nu}. \label{sources}
\ee

A different view on (\ref{sources}) comes from a matrix model-like formulation of electromagnetism, 
derived as follows: (\ref{HemF}) implies
$\ddot{q}_i=\frac{1}{2i}([\dot{q}_i,\dot{q}_j]\dot{q}_j+\dot{q}_j[\dot{q}_i,\dot{q}_j])$,
or, working with $p_i=\dot{q}_i$,
\be
\dot{p}_i
=\frac{1}{2i}(p_i p^2_j-p^2_jp_i)
=\frac{1}{2i}[[p_i,p_j],p_j]_+
=\frac{1}{2i}[[p_i,p_j]_+,p_j], \label{mmF}
\ee
where $[p_i,p_j]_+\equiv p_i p_j+p_j p_i$. 
The first order differential equations (\ref{mmF}) describe the evolution of $p_i$,
and thus implicitely the one of $i F_{ij}=[p_i,p_j]$, suggesting that
the equations with sources (\ref{sources}) are not needed. However, (\ref{mmF})
has an infinite number of solutions, and to single out one extra constraints are needed. 
They are provided precisely by Eqs. (\ref{sources}). 

Finally, we extend the former procedure to non-Abelian connections. 
For this purpose we allow multi-component wave functions, 
assigning to the wave-function an internal index $a=1,2,\dots,n$,
\be
[p_{\mu},p_{\nu}]^{ab}\Psi_{b}=iF_{\mu\nu}^{ab}\Psi_b.  \label{cr_ym}
\ee
The commutation relations (\ref{cr_ym}) can be represented by 
\be
p_{\mu}^{ab}=-i\partial_{\mu}\delta^{ab}-A_{\mu}^s T_s^{ab}, \label{p_ym}
\ee
where $T_s^{ab}$ are matrices that span the group space to which $F_{\mu\nu}^{ab}$ belongs
as far as its action related to the indices $a$ and $b$ is concerned. Eqs. (\ref{cr_ym},\ref{p_ym}) require that
\be
F_{\mu\nu}^{ab}=\partial_{\mu} A_{\nu}^s T_s^{ab}-\partial_{\nu} A_{\mu}^s T_s^{ab}
+A_{\mu}^s A_{\nu}^t [T_s,T_s]^{ab}
\ee
and the Jacobi identity implies that $F$ does not depend on $p$, and that
$\partial_{\rho}F_{\mu\nu}^{ab}+\partial_{\nu}F_{\rho\mu}^{ab}+\partial_{\mu}F_{\nu\rho}^{ab}=0$.
The operatorial equations of motion for a test particle 
are derived like in the abelian case, and read
\be
\ddot{q}_{\mu}^{ab}=\frac{1}{2}(F_{\mu\nu}^{ac}p_{\nu}^{cb}+p_{\nu}^{ac}F_{\mu\nu}^{cb}).
\ee

\subsection{Gravity}

Another possibility for the $[p,p]$ commutator is to have the wave function carry a space-time index. 
The commutation relation is then 
\be
[p_{\mu},p_{\nu}]\Psi_{\rho}=iR^{\sigma}_{\mu\nu\rho}\Psi_{\sigma}. \label{gfromp}
\ee
Obviously, $R^{\sigma}_{\mu\nu\rho}$ is intended to be the Riemann tensor.
One may realize (\ref{gfromp}) via $p_{\mu}=iD_{\mu}$, where
\be
D_{\mu}\Psi_{\nu}=
(\partial_{\mu}\delta^{\sigma}_{\nu}+i\Gamma^{\sigma}_{\mu\nu})\Psi_{\sigma}. \label{D}
\ee
(We assume that the covariant derivative is Leibniz.) Eqs. (\ref{gfromp},\ref{D}) lead to 
\be
R^{\kappa}_{\lambda\mu\nu}=\partial_{\mu}\Gamma^{\kappa}_{\lambda\nu}-\partial_{\nu}\Gamma^{\kappa}_{\lambda\mu}+
i\Gamma^{\kappa}_{\sigma\mu}\Gamma^{\sigma}_{\lambda\nu}-i\Gamma^{\kappa}_{\sigma\nu}\Gamma^{\sigma}_{\lambda\mu}
\label{R_im}
\ee 
with $R$ satisfying all the usual symmetry properties (including Bianchi, again derived from Jacobi).
However, due to the extra $i$, the $R$ in (\ref{R_im}) is complex, 
not the real (in both senses) $R$ we are after.
The mismatch is due to the real gravity connection appearing at the classical level, via
$\partial \rightarrow \partial+\Gamma $, without the extra $i$ the approach (\ref{gfromp}) 
introduces
\footnote{Weyl encountered the opposite problem in his first attempt 
to derive electromagnetism from gauge invariance \cite{weyl_1918}. 
His classical approach, based on a conformal extension of general relativity, lacked a crucial $i$. 
In trying to derive gravity from the quantum-mechanical 
commutation relations (\ref{gfromp}) one extra $i$ appears.}. 
We conclude that $[p,p]$ commutators cannot generate Einstein's connection.

There is however a different, quite natural, way to introduce the effect of a gravitational background, 
by postulating
\be
[q^{\mu},p^{\nu}]=i g^{\mu \nu}(q^{\sigma}) \qquad [q^{\mu},q^{\nu}]=0 \qquad [p_{\mu},p_{\nu}]=0,
\label{metric}
\ee
with $g^{\mu \nu}$ symmetric 
\footnote{A nonsymmetric $g$ can also be used, by taking
$p^k\equiv \frac{1}{2}(g^{km}+g^{mk})p_m$ in the following.}
and nondegenerate, and $p_{\kappa}\equiv g_{\kappa\mu}(q^{\sigma})p^{\mu}$. 
The inverse of $g^{\mu \nu}$ is $g_{\mu \nu}$: $g^{\mu \nu}(q^{\sigma})g_{\nu\kappa}(q^k)=\delta^{\mu}_{\kappa}$.
Now $q^{\mu},p^{\nu}$ depend on the proper time $\tau$, and $\frac{do}{d\tau}\equiv \dot{o}$.
The Hamiltonian is taken to be
\be
H=\frac{1}{2}p_{\mu}p^{\mu}=\frac{1}{2}p_{\mu}p_{\nu}g^{\mu \nu}(q^{\rho}),
\label{hamiltonian}
\ee
with a suitable ordering implied.
The main point is to distinguish between upper and lower indices. Because
$
[q^{\mu},p_{\nu}]=\delta^{\mu}_{\nu},
$
$[\cdot,p_{\nu}]$ can be used as a derivative operator on functions $f(q^{\mu})$.
Since the nonconstancy of $g^{\mu\nu}(q^{\kappa})$ may lead to operator ordering issues
we start classically, with the following Poisson brackets:
\be
\{q^{\mu},p^{\nu}\}=g^{\mu\nu}(q^{\kappa}),  \qquad \{q^{\mu},q^{\nu}\}=0,  \qquad \{p_{\mu},p_{\nu}\}=0,  
\qquad \label{PBg}
\ee
($q$ and $p$ are now classical variables, not operators). $\{q^{\mu},p_{\nu}\}=\delta^{\mu}_{\nu}$ implies 
\be
\{f(q^{\mu}),p_{\nu}\}=\frac{\partial f}{\partial q^{\nu}}\equiv \partial_{\nu} f \label{derivative1}.
\ee
Half of the Hamilton equations give
\be
\dot{q}^{\kappa}=\{q^{\kappa}, H\}=\{q^{\kappa},\frac{1}{2}p_{\mu}p_{\nu} g^{\mu\nu}(q^{\sigma}) \}
=p^{\kappa}, \label{qdot}
\ee
whereas the second half of them reads
\be
\ddot{q}^{\kappa}=\{p^{\kappa}, H\}=\{p_{\lambda} g^{\lambda\kappa}(q^{\sigma}),
\frac{1}{2}p_{\mu}p_{\nu} g^{\mu\nu}(q^{\sigma}) \}.
\ee
Using the relation  
$\partial_{\kappa} g^{\mu\nu}=-(\partial_{\kappa} g_{\sigma\tau}) g^{\mu\sigma} g^{\nu\tau}$ 
and Eqs. (\ref{derivative1},\ref{qdot}) one obtains
\be
\ddot{q}^{\kappa}+\Gamma^{\kappa}_{\mu\nu}\dot{q}^{\mu}\dot{q}^{\nu}=0.  \label{emg}
\ee
Above,  $\Gamma$ is the usual Christoffel connection,
\be
\Gamma^{\kappa}_{\mu\nu}=
\frac{1}{2}g^{\kappa\lambda}(\partial_{\nu} g_{\mu\lambda}+\partial_{\mu} g_{\nu\lambda}-\partial_{\lambda} g_{\mu\nu}). 
\label{connection}
\ee
Eqs. (\ref{emg},\ref{connection}) describe the motion of a point-particle in a 
gravitational background, represented by the metric $g^{\kappa\mu}(q)$.

At this point, the Riemann tensor can be defined through
\be
R^{\kappa}_{\lambda\mu\nu}=\partial_{\mu}\Gamma^{\kappa}_{\lambda\nu}-\partial_{\nu}\Gamma^{\kappa}_{\lambda\mu}+
\Gamma^{\kappa}_{\sigma\mu}\Gamma^{\sigma}_{\lambda\nu}-\Gamma^{\kappa}_{\sigma\nu}\Gamma^{\sigma}_{\lambda\mu}
\label{riemann}
\ee
and all its symmetries follow in the usual way. The same applies to gauge invariance. 
In contrast to electromagnetism and the unsuccessful attempt (\ref{gfromp}), the symmetries
and the Bianchi identity of (\ref{riemann}) do not follow from the Jacobi identities,
which are in fact identically satisfied, as straightforward calculation shows. 
Thus no constraints are imposed on the metric, which is a genuine Einstein gravity background.

To proceed to the quantum-mechanical level, one has to prescribe an ordering, 
be it left, right, symmetric, or whatever, of the operators $q$ and $p$.
Any of those orderings satisfies 
\be
[p_{\kappa}, <f(q^{\mu})>]=-i  \partial_{\kappa} <f> \equiv -i  <\partial_{\kappa} <f>>,\label{ordprop}
\ee
where $<\dots>$ denotes the operation of taking a given ordering.
Eq. (\ref{ordprop}) and the commutativity of operators under ordering symbols
allow one to adapt step by step the derivation (\ref{PBg}--\ref{emg}) to the operatorial case.
The Schr\"odinger equation $H(q,p)\Psi=-i\partial_t\Psi$ is a more delicate issue, being ordering dependent.
The kinetic part of $H$ gives, in addition to the Laplace-Beltrami operator, an ordering dependent correction.
This well-known issue is discussed at length from the usual point of view in \cite{ordering}.


Considering simultaneously nontrivial $[p,p]$ and $[q,p]$ commutators 
only amounts to a superposition of electromagnetic and gravitational backgrounds. 
The constraints imposed by the Jacobi identity remain those of a flat background
(i.e. independence of $F$ from $p$ and Bianchi for $F(q)$).

On the other hand, one can allow the metric to depend on both $q$ and $p$, $g(q,p)$, 
This leads to the same Jacobi constraints as above (trivially satisfied only when $F$ is zero), 
but to equations of motion more complicated than (\ref{emg}). 
The first Hamilton equation,
\be
\dot{q}^{\kappa}=\{q^{\kappa},\frac{1}{2}p_{\mu}p_{\nu} g^{\mu\nu}(q,p)\}
=p^{\kappa}-\frac{1}{2}p_{\mu}p_{\nu} \partial_{p_{\kappa}}g^{\mu\nu}(q,p), \label{h1_gqp}
\ee
already shows that it becomes difficult even to express $p^{\kappa}$ as a function of $\dot{q}^{\mu}$.
The second Hamilton equation is quite complicated, especially when $F\neq 0$, and is not reproduced here. 
One particularly simple case stands out however, namely $g(p)$ and $F=0$. 
Then $\dot{q^{\kappa}}$ and $H$ in (\ref{h1_gqp}) are exclusively functions of $p_{\mu}$,
$\ddot{q^{\kappa}}=\{\dot{q}^{\kappa}(p_{\mu}),H(p_{\nu})\}=0$, and the test particle has flat space motion
This remains true upon introduction of noncommutativity, $\theta\neq 0$, 
but is completely spoiled if $F\neq 0$.



We generated the three  known types of gauge backgrounds via $[p,p]$ and $[q,p]$ commutators. 
We proceed with nontrivial $[q,q]$ commutators.


\section{Noncommutative space}


\subsection{Electromagnetism and noncommutativity}


Consider first a generalized electromagnetic background $F(q,p)$, living on a space
with noncommutativity field $\theta(q,p)$, and flat metric $g_{ij}=\delta_{ij}$: 
\be
[q^i,q^j]=i\theta^{ij}(q,p)  \qquad [q^i,p^j]=i\delta^{ij}    \qquad [p_i,p_j]=iF_{ij}(q,p).   \label{PB_FT} 
\ee
The relativistic generalization $\delta^{ij}\rightarrow \eta^{\mu\nu}$ is straightforward. 
To avoid explicit referral to ordering issues, we work with Poisson brackets, i.e. $-i[,]\rightarrow \{,\}$.

If the Hamiltonian depends only on the momenta, $H(p)$, 
the equations of motion are not modified by a nonzero $\theta^{ij}$.
The Jacobi identities change however, no matter what the Hamiltonian is; they read
\be
\{q^k,F_{ij}\}=\frac{\partial F_{ij}}{\partial p_k}-\frac{\partial F_{ij}}{\partial q^m}\theta^{mk}=0   \label{j1}
\ee
\be
\{\theta^{ij},p_k\}=\frac{\partial \theta^{ij}}{\partial q^k}+\frac{\partial \theta^{ij}}{\partial p_m}F_{mk}=0  
\label{j2}
\ee
\be
\{F_{ij},p_k\}+cyclic=
(\frac{\partial F_{ij}}{\partial q^k}+\frac{\partial F_{ij}}{\partial p_m}F_{mk})+cyclic =0 \label{jc1}
\ee
\be
\{q^k,\theta^{ij}\}+cyclic=
(\frac{\partial \theta^{ij}}{\partial p_k}-\frac{\partial \theta^{ij}}{\partial q^m}\theta^{mk})+cyclic =0. \label{jc2}
\ee
The Jacobi identities ensure the invariance of the commutation relations
under time evolution, for a generic Hamiltonian $H(p,q)$. Explicitely:
\be
\{\{q^m,p_n\}-\delta^m_n,H\}=-\partial_{p_s}H\{F_{ns},q^m\}+\partial_{s}H\{\theta^{ms},p_n\},\label{ev1}
\ee
\be
\{\{q^m,q^n\}-\theta^{mn},H\}=-\partial_{p_s}H\{\theta^{mn},p_s\}-
\partial_{s}H(\{\theta^{ns},q^m\}+cyclic),\label{ev2}
\ee
\be
\{\{p_m,p_n\}-F_{mn},H\}=-\partial_{s}H\{F_{mn},q^s\}
+\partial_{p_s}H(\{F_{mn},p_s\}+cyclic),\label{ev3}
\ee
and  Eqs. (\ref{j1}--\ref{jc2}) ensure the right hand side in (\ref{ev1}--\ref{ev2}) to be zero.
They also restrict $F$ and $\theta$ to 
one of the following forms:
\begin{itemize}
\item $F$ and $\theta $ both constant
\item $\theta $  constant, and $F(q,p)$  constrained by (\ref{j1},\ref{jc1})
\item $F $  constant, and $\theta(q,p)$  constrained by(\ref{j2},\ref{jc2})
\item $F(q,p)$ and $\theta(q,p)$, both constrained by (\ref{j1}--\ref{jc2})
\end{itemize}
In particular, (\ref{j1}) forbids an electromagnetic field strength to depend only on the coordinates,
$F_{ij}(q)$, if $\theta^{ij}\neq 0$.
We proceed to discuss all the above situations in turn; in most cases $\theta$ and $F$ are related.



\vskip 0.1cm

{\boldmath $\theta$ {\bf and} $F$ {\bf constant}},
or noncommutative quantum mechanics \cite{ncqm}, the simplest case. It is not our main interest here.

\vskip 0.1cm

{\boldmath $F$ {\bf constant and} $\theta(q,p)$}.
Eq. (\ref{j2}) constrains the functional dependence of $\theta$ to be of the form
$\theta(\bar{p}_m)=\theta(p_m-F_{mn}q^n)$. To satisfy  (\ref{jc2}) also,
one can either block-diagonalize $\theta$, or require 
$\partial_{\bar{p}_s}\theta^{ij}(\bar{p})[\delta_s^k+F_{sm}\theta^{sk}(\bar{p})]+cyclic =0$. 
The equations of motion do not change if $H(p)$, they remain those of a classical particle in a constant electromagnetic field. 
The Schr\"odinger equation in (2+1)-dimensions \cite{pi} confirms that: 
$\psi(q^1,p_2)$ satisfies the same equation as when $\theta\equiv 0$.
The only difference is that objects like $\psi(q^1,q^2)$ and $\psi(p_1,p_2)$ are not definable anymore.

Two limiting cases exist:

1. {\em $F=0$  and $\theta(p)$}, which implies a 'p-Bianchi identity' for $\theta$, 
\be
\partial_{p_k}\theta^{ij}(p)+\partial_{p_i}\theta^{jk}(p)+\partial_{p_j}\theta^{ki}(p)=0. \label{p-Bianchi}
\ee
$\theta(p)$ plays the role of a magnetic field in momentum space \cite{pi}. To make the 
duality $q\leftrightarrow -p$ fully manifest one has however to introduce a harmonic potential, $V\sim q^2$. 
Again, if $\partial_q H(p,q)= 0$, $\theta$ has no effect on the equations of motion, 
which remain those of a free particle.

2. {\em $F^{-1}\rightarrow 0$  and $\theta\rightarrow \theta(q)$}.
If one allows a potential term also, $H(q,p)=p^2/2+V(q)$, given that now $F\rightarrow \infty$, 
one gets $\ddot{q}_k\simeq F_{km}(\dot{q})(\dot{q}_m-\theta^{ms}\partial_s V)$.

\vskip 0.15cm

{\boldmath $F(q,p)$ {\bf and} $\theta$ {\bf constant.}} {\bf Experimental signals.} 

If $\theta$ is constant, Eq. (\ref{j1}) requires that $F=F(\bar{q}^m)=F(q^m+\theta^{mn}p_n)$.
To satisfy (\ref{jc1}) one can either partially block-diagonalize $F$ and $\theta$ 
such that they couple only pairs of directions,
or just look for a solution of (\ref{jc1}) for $F(\bar{q}^m)$. (\ref{jc1}) is automatically
satisfied in two dimensions (2D), where our main interest will be.
For small $\theta$ one obtains, for a particle of mass $m$,
\be
m\ddot{q}^m=F_{mn}(q^s+\theta^{st}p_t)\dot{q}^n\simeq 
F_{mn}(q^s)\dot{q}^n+m\partial_s F_{mn}(q) \theta^{st}\dot{q}^t\dot{q}^n \label{test1}.
\ee
We have thus the superposition of a usual electromagnetic background, linear in velocities,
and of a force quadratic in the velocities, $\partial_s F(q)_{mn} \theta^{st}\dot{q}^t\dot{q}_n$.
$\gamma^s_{tn}\equiv \partial_s F(q)_{mn} \theta^{st}$ simulates a gravitational connection.
This interpretation is however valid only in one reference frame, since 
$\gamma^s_{tn}$ does not behave like a Christoffel symbol under generalized coordinate transformations.

It is important that a nonzero noncommutativity, $\theta\neq 0$, {\em requires} $F(q,p)$.
Noncommutativity might consequently be detected through the effects of the additional term in (\ref{test1}).
A simple set-up \cite{bound1} actually provides a good bound for $\theta$.
Consider  motion in a plane labelled by coordinates $q_1=x$, $q_2=y$,
and assume a magnetic field  of spatial dependence $F(x,y)=B e(y)$, with $e(y)$ the Heaviside step function,
the integral of Dirac's delta function, $\frac{de(y)}{dy}=\delta(y)$ (figure 1a).

\setlength{\unitlength}{1cm}
\begin{picture}(12,5)
\put(1.7,0.7){figure 1a}
\put(.5,3){\line(1,0){4}}
\put(.8,4){{\large B}}
\put(1.5,2.6){\small{P}}
\put(3.1,2.6){\small{P'}}
\put(1.5,3){\circle*{.1}}
\put(3.5,3){\circle*{.1}}
\qbezier(1.5,3)(2.5,4)(3.5,3)
\put(1.5,3){\line(-1,-2){.8}}
\put(3.5,3){\line(3,-4){1}}
\put(0.3,2.2){$\vec{v}_0$}
\put(1.8,3.6){$\vec{v}'$}
\put(4.3,2.5){$\vec{v}''$}
\put(4.8,1.5){$\vec{v}$}
\thicklines
\put(1.5,3){\vector(1,1){.6}}
\put(3.5,3){\vector(1,-1){.6}}
\put(.75,2){\vector(1,2){.4}}
\put(4.09,2.2){\vector(3,-4){.5}}
\thinlines
\put(8.3,0.7){figure 1b}
\put(7.2,4){{\large B}}
\put(8,2.6){\small{P}}
\put(9.9,3.4){\small{P'}}
\put(8,3){\circle*{.1}}
\put(10,3.2){\circle*{.1}}
\put(6.8,2.2){$\vec{v}_0$}
\put(8.25,3.6){$\vec{v}'$}
\put(10.5,3){$\vec{v}''=\vec{v}$}
\put(7,3){\line(1,0){2.8}}
\put(9.8,3){\line(1,1){.5}}
\qbezier(8,3)(9,4)(10,3.2)
\put(8,3){\line(-1,-2){.8}}
\put(10,3.2){\line(1,-1){1}}
\thicklines
\put(8,3){\vector(1,1){.6}}
\put(10,3.2){\vector(1,-1){.6}}
\put(7.25,2){\vector(1,2){.4}}
\end{picture}

\noindent
The particle, with initial velocity $\vec{v}_0$, penetrates at point P in the half-plane with nonvanishing
magnetic field. There, its velocity will change to $\vec{v}'$, with which it starts the usual circular motion
in the field $B$. While exiting the superior half-plane, its velocity will change to a final $\vec{v}''$.  
Eq. (\ref{test1}) leads to 
\be
m\ddot{x}-F(y)\dot{y}=-m\theta B\delta(y)\dot{x}\dot{y}, \qquad
m\ddot{y}+F(y)\dot{x}=m\theta B\delta(y)\dot{x}^2.
\ee 
Denoting $\dot{x}=v_x,\dot{y}=v_y$, and using $\delta(y)=\delta(v_y t)=\frac{\delta(t)}{|v_y|}$,
one has
\be
m\ddot{x}-F(y)v_y=-m\theta B\delta(t)\dot{x}\frac{v_y}{|v_y|}, \qquad
m\ddot{y}+F(y)v_x=m\theta B\delta(t)\frac{v_x^2}{|v_y|}, \label{shift}
\ee 
and integrating from $t=-\epsilon$ to $t=\epsilon$ (and $\epsilon\rightarrow 0$) one obtains
\be
v'_x=v_x(1-\theta B),\qquad v'_y=v_y(1+\theta B\frac{v^2_x}{v^2_y}).
\ee
An instantaneous velocity change - and no instantaneous displacement - 
appears at the point at which the particle enters the magnetic field. 
The change in kinetic energy is of order
$\theta^2$, $(v'_x)^2+(v'_y)^2=v^2_x+v^2_y+\theta^2B^2(v^2_x+v^4_x/v^2_y)$.
The opposite mechanism works while the particle exits the magnetic field region.
This renders the correction quadratic in $\theta$, i.e. quite small.
To avoid that, figure 1b proposes a slightly modified configuration, which allows the particle
to exit the magnetic field along the gradient. In this case no unconventional velocity shift
occurs at P', and the particle maintains the $O(\theta)$ effects of (\ref{shift}). 

It is important that the noncommutative correction (\ref{shift}) depends also on the ratio $\frac{v^2_x}{v^2_y}$;
this allows us to put ourselves in a favorable situation.
Assuming $\frac{v_x}{v_y}\sim 10^4$, $B\sim 1 T$, a proton of mass $m_p$ and charge $1.6\times 10^{-19}C$,
and a one percent sensitivity to velocity changes, one obtains $\sqrt{\theta}\sim10^{-14}m$ \cite{bound1}.
If one constructs a system of magnets keeping the proton under periodic motion, one further enhances the bound to
\be
\sqrt{\theta}\sim 10^{-14-n}m. \label{bound1}
\ee 
where $10^{2n}$ is the number of revolutions performed by the particle, without any noncommutative effect appearing.
Accurate control for $n=5$ seems an easy task, implying that one can probe noncommutativity in the $TeV$ region with
a classical, low-energy and cheap experiment!

One can also study the noncommutative Aharonov-Bohm effect using the above considerations.
A path integral formalism \cite{pi} is convenient, and the final result is that the phase shift
produced by a magnetic field $B$ inside a thin solenoid does depend on $B(1+\theta B)$, instead of $B$. 
The bound  on $\theta$ is weaker then (\ref{bound1}). 

Using the relativistic formulation, one can also use the Hydrogen atom to put bounds on $\theta$.
For, if $\theta^{12}=\theta\neq 0$,
then $F_{0i}(x+\theta \dot{y},y-\theta\dot{x},z)$: the Coulomb potential displays a 
velocity-dependent and anisotropic correction
$V \sim [x^2+y^2+z^2+2\theta(x\dot{y}-y\dot{x})]^{-\frac{1}{2}}\simeq\frac{1}{r}
(1-\frac{\theta(x\dot{y}-y\dot{x})}{r^2})$. 
As the present work focuses on the theoretical structure implied by nonconstant backgrounds,
the effects of the above potential will be discussed separately. 

Finally, let us mention the two limiting cases: 

1. {\em $\theta= 0$  and $F(q)$}, which is usual electromagnetism.

2. {\em $\theta^{-1}\rightarrow 0$  and $F\rightarrow F(p)$}, which has the following dynamical content:
If $H(p)=p^2/2$, get $\ddot{q}_k=F_{km}(\dot{q})\dot{q}_m$, i.e. first order differential equation in $\dot{q}$.
If $H(q,p)=p^2/2+V(q)$, due to $\theta\rightarrow \infty$, the potential term will dominate the dynamics;
in a first approximation $\ddot{q}^k\simeq \theta^{ks}\theta^{lm}\partial_l\partial_sV\partial_mV$. 


\vskip 0.15cm

{\boldmath $F(q,p)$ {\bf and} $\theta(q,p)$}. {\bf Dimensional reduction}.

We proceed with the case in which both $\theta$ and $F$ are nonconstant.
The Jacobi identities imply that they should depend on both $q$'s and $p$'s.
Since handling the generic Jacobi identities is somehow messy,
we work in the 2D case, $q_1,q_2,p_1,p_2$ depending on $t$, $F_{12}\equiv F$, $\theta_{12}\equiv\theta$.
Then, (\ref{jc1},\ref{jc2}) are identically satisfied, whereas (\ref{j1},\ref{j2}) read:
\be
F\partial_{p_2}\theta=\partial_{1}\theta \qquad  F\partial_{p_1}\theta=-\partial_2\theta \label{j2dt}
\ee
\be
\partial_{p_2}F=\theta\partial_1{F}\qquad  \partial_{p_1}F=-\theta\partial_2{F}.\label{j2df}
\ee
If we assume $\theta$ small and slowly varying, we may try the solution (\ref{test1}),
$F=F(q^i+\theta^{ij}(q,p)p_j)$, $i,j=1,2$. Expanding in powers of $\theta$ we arrive  at the constraint 
$\theta\sim F^{-1}$.
Interestingly, if we assume that $F$ depends only on $q_1$ and $p_2$, and try to solve (\ref{j2dt},\ref{j2df})
by separation of variables, we end-up again with $F=\theta^{-1}$. 
The same happens if we assume $F(q_2,p_1)$, the solution obtained by separation of variables being
\be
F(q^2,p_1)=\frac{p_1+c_1}{q^2+c^2}=\frac{1}{\theta(q^2,p_1)}.
\ee
$c_1$ and $c^2$ are two arbitrary constants. Actually, a more general solution of 
(\ref{j2dt},\ref{j2df}) can be found, of the form
\be
F(q^1,q^2,p_1,p_2)=\frac{p_1+p_2+c_1}{q^2-q^1+c^2}=\frac{1}{\theta(q^1,q^2,p_1,p_2)}.
\ee 
Also, if one assumes $\theta=F^n$, one finds solutions only if $n=-1$.
If $F=\theta^{-1}$, half of the equations in (\ref{j2dt},\ref{j2df}) trivialize.
The above solutions and remarks suggest that, in general,
the Jacobi identities (\ref{j2dt},\ref{j2df}) imply that, if $F$ and $\theta$ are both nonconstant,
they are related through 
\be
F=\frac{1}{\theta}. \label{dimred}
\ee
This is an important statement, as it will be shown below that (\ref{dimred}) implies
a 'collapse' of the phase space from four dimensions to two.
This dimensional reduction is independent of the dynamics; it takes place for any Hamiltonian.
We do not have a general proof that {\em all} the solutions of (\ref{j2dt},\ref{j2df})
imply (\ref{dimred}), but we further support this conjecture by noticing that, due to the 
$F \longleftrightarrow \theta^{-1}$ symmetry of (\ref{j2dt},\ref{j2df}), 
$F$ and $1/\theta$ satisfy the same equations
\be
+f(\partial_1 \partial_{p_2} f \partial_{p_2} f-\partial_1 f\partial_{p_2}^2 f)=
\partial_1^2 f \partial_{p_2} f-\partial_1 f \partial_1\partial_{p_2} f  
\ee
\be
-f(\partial_2 \partial_{p_1} f \partial_{p_1} f-\partial_2 f\partial_{p_1}^2 f)=
\partial_2^2 f \partial_{p_1} f-\partial_2 f \partial_2\partial_{p_1} f   
\ee
where either $f=F$ or $f=1/\theta$. 
(The additional $q_1, p_2 \longleftrightarrow -q_2,p_1$ symmetry of the system is also evident.)
Moreover, $F$ and $1/\theta$ are connected through the equations in (\ref{j2dt},\ref{j2df}). 
It seems that there is little space left for additional solutions not satisfying (\ref{dimred}).

In consequence, it appears that three types of noncommutative space can exist:
1. $\theta$ and $F$ both constant; 
2. Either $\theta$ or $F$ constant;
3. None constant, with dimensional reduction taking place spontaneously.





We will now show that $F=\frac{1}{\theta}$ arises if and only if the $q$'s and $p$'s are not independent,
i.e. if dimensional reduction takes place. 
For generality, we work in an arbitrary number of dimensions.
Whenever we revert to 2D parlance, we will write as before $\theta\equiv \theta^{12}, F\equiv F_{12}$.
First, assume there exists a relation between the $q^i$'s and the $p_j$'s, say,
\be
q^n=g^n(p_m) \qquad p_m=f_m(q^s)=(g^{-1})_m(q^s). \label{dependence}
\ee
Asking consistency of the commutation relations, one obtains
\be
\frac{\partial q^m}{\partial p_s}=-\theta^{ms} \qquad 
\frac{\partial p_k}{\partial q^n}=F_{kn}.  \label{in_fact}
\ee 
This follows, for instance, from $\{f(q(p)),q^s\}=\frac{\delta f}{\delta q^m}\theta^{ms}
=-\frac{\delta f}{\delta q^m}\frac{\partial q^m}{\partial p_s}$.
Since $\frac{\partial q^m}{\partial p_k}\frac{\partial p_k}{\partial q^n}=\delta^m_n$, 
it follows that 
\be
\theta^{mk}F_{kn}=-\delta^m_n, \label{maestru}
\ee
or $F_{12}\equiv F=\theta^{-1}\equiv \theta_{12}^{-1}$ in 2D parlance. 
(\ref{dimred}) is thus enforced by (\ref{dependence}) and (\ref{PB_FT}).
An alternative derivation uses the following chain of equalities,
$\delta^m_n=\{q^m,p_n\}=\{g^m(p),p_n\}=\frac{\partial q^m}{\partial p_s}F_{sn}=-\theta^{mk}F_{kn}$. 
Eq. (\ref{maestru}) 
is fully consistent with the commutation relations:
\be
\{q^m,q^n\}=\{g^m(p),g^n(p)\}=
\frac{\partial g^m}{\partial p_s}F_{st}\frac{\partial g^n}{\partial p_t}=
\theta^{ms}F_{st}\theta^{nt}=\theta^{mn}.
\ee
Conversely, (\ref{maestru}) implies (\ref{dependence}), with the relationship
between the $q$'s and the $p$'s actually taking the more precise from
\be
q^m=-\theta^{mn}(q,p)p_n+cst.   \qquad  \forall H(p,q) \label{explicit}
\ee
To prove this, first notice that (\ref{maestru}) and 
the equations of motion imply that
\be
\dot{q}^m+\theta^{mn}\dot{p}_n=\{q^m+\theta^{mn} p_n,H\}=0, \quad \forall H(p,q).\label{variations}
\ee
This already shows that dimensional reduction occurs,
since the variations of the $q$'s and the $p$'s are related. 
Eq. (\ref{explicit}) follows from Eq. (\ref{variations})
once $\dot{\theta}=0$. 
To prove that, first observe that Jacobi implies $\{F,q^i\}=0$ and $\{\theta,p_j\}=0$, hence
\be
\dot{F}=\{F,p_i\}\partial_{p_i} H \qquad \dot{\theta}=\{\theta,q^j\}\partial_{q^j} H.\label{tdot}
\ee
However, if  $\theta F =1$,  
it happens that both $\dot{\theta}=0$ and $\dot{F}=0$, $\forall H(p,q)$. This is so because
if $\theta$ is a function of $F$ or viceversa, then the nonzero terms in (\ref{tdot}) become zero, due to
$\{\theta,q^i\}=\frac{\partial \theta}{\partial F}\{F,q^i\}=0$,
$\{F,p_j\}=\frac{\partial F}{\partial \theta}\{\theta,p_j\}=0$.
Thus $F(q,p)$ and $\theta(q,p)$ are constants of motion, for instance functions of the Hamiltonian.
If $\dot{\theta}=0$, then $\dot{q}=\theta\dot{p}$ {\em implies} $q=\theta p$, and viceversa, 
and (\ref{explicit}) is proved. 
We note in passing that if $H(p)$, $\dot{\theta}=0$ immediately, due to Jacobi, cf. (\ref{tdot}),
and $\theta(q,p)$ is a constant of motion.

One may ask now: under what conditions are (\ref{explicit}) and (\ref{in_fact}) compatible?
It turns out (differentiating with respect to $p$) that those conditions are precisely the Jacobi identities,
now meaning that the total variation of $F$ or $\theta$ with respect to one set of coordinates ($q$ xor $p$) is zero.
That the Jacobi identity $\{\theta^{mn},p_l\}=0$ means zero  total variation of $\theta$ in reduced space
can be seen from the following sequence of equalities:
\be
\{\theta^{mn},p_l\}=\frac{\partial \theta^{mn}}{\partial q^l}+F_{sl}\frac{\partial \theta^{mn}}{\partial p_s}\\
=F_{sl}\left (\frac{\partial \theta^{mn}}{\partial p_s}+
\frac{\partial \theta^{mn}}{\partial q^r}\frac{\partial q^r}{\partial p_s}\right )
=F_{sl}\frac{\delta \theta^{mn}}{\delta p_s}=\frac{\delta \theta^{mn}}{\delta q^l}
\ee
in which we used (\ref{in_fact}) repeatedly. Similarly, $\{F_{mn},q^l\}=0$ implies
$\frac{\delta F_{mn}}{\delta q^t}=F_{st}\frac{\delta F_{mn}}{\delta p_s}=0$.
This is satisfied either by constant $F$ and $\theta$, or by such a dependencies which cancels
when $q=q(p)$. 
In 2D, such examples are $F_{12}(p_1,q_2)=\theta^{-1}_{12}=\frac{p_1}{q_2}$; 
$F_{12}(p_2,q_1)=\theta^{-1}_{12}=-\frac{p_2}{q_1}$,
or a more general one, $F_{12}(p_1,p_2,q_1,q_2)=\theta^{-1}_{12}=\frac{p_1+p_2}{q_2-q_1}$.
It is easy to check that any of those $F$'s satisfies $\{F,q^l\}=0$, once $F=\theta^{-1}$.

For $\theta$ and $F$ constant in  (\ref{maestru}), the dimensional reduction 
occuring when $F=\theta^{-1}$ has long been known in a slightly different form \cite{peierls}.
Here we generalized it to the case of nonconstant backgrounds.

\subsection{Gravity and noncommutativity}


Consider for simplicity the case in which the metric depends only on the coordinates,
$g(q)$, 
and the electromagnetic-like background vanishes, $F=0$:
\be
\{q^i,p_j\}=\delta^i_j \quad \{q^i,q^j\}=\theta^{ij}\quad \{p_i,p_j\}=0,
\ee
and $p^i=g^{ij}(q^s)p_j$, $2H=p_i p^i$ [going to Minkowskian signature is straightforward:
$\delta_{ij}\rightarrow \eta_{\mu\nu}$, $q^i\rightarrow q^{\mu}=(q^0,q^i)$, 
$p_i\rightarrow p_{\mu}=(p_0,p_i)$, $t\rightarrow \tau$]. 
If $\theta\neq 0$, the bracket $\{q^i,p^j\}$ is not symmetric anymore, and one  
gets also a nonzero  $\{p^i,p^j\}$ bracket.
The Jacobi identities read the same as in flat space, 
but the equations of motion change.
First, we have
\be
\dot{q}^k=p^k+\frac{1}{2}p_ip_j \partial_s g^{ij}\theta^{ks},
\ee
and
\be
\ddot{q}^k=\{g^{kj}p_j+\frac{1}{2}p_ip_j \partial_s g^{ij}\theta^{ks},\frac{1}{2}p_mp_n  g^{mn}\}.
\ee
For small $\theta$, one obtains 
\be
p^k\simeq \dot{q}^k+\frac{1}{2}p^mp^n \theta^{ks} \partial_s g_{mn}
\ee
and in the end one gets the following equations of motion
\be
\ddot{q}^k= -\Gamma^k_{mn}\dot{q}^m\dot{q}^n+ \Delta^k_{mnp}\dot{q}^m\dot{q}^n\dot{q}^p, \label{test2}
\ee
with the correction to the commutative motion given by
\be
\Delta^k_{mnp}=\Gamma^l_{np}\theta^{ks}\partial_s g_{lm}-\Gamma^k_{pl}\theta^{ls}\partial_s g_{mn}
+\frac{1}{2}g^{lk}\{g_{lm},g_{np}\}-\theta^{kl}\frac{1}{2}\partial_l \partial_pg_{mn}.
\ee
It is understood that $\{g_{lm},g_{np}\}=\partial_s g_{lm}\theta^{st}\partial_t g_{np}$.
The correction is proportional to the first power of $\theta$ only for small $\theta$.
It then happens that 
while $F$ and $g$ introduce in the equations of motion terms linear, 
respectively quadratic, in the velocities $\dot{q}^m$,
$\theta$ brings in terms {\em cubic} in the velocities. 
The equation of motion for a test particle (\ref{test2}) can be used to put a stringent bound
on $\theta$; this will be discussed in detail elsewhere.

\subsection{Significance of $\theta$}

We saw that $F$ is related to the field-strenght of a gauge field, 
and $g$ to the metric of a gravitational theory. 
One may then ask: what does $\theta$ correspond to?
%
%

Although several points of view are possible, we present here only the simplest one.
One can first go to a higher dimensional space,
by promoting the phase-space $\{q,p\}$ to an ordinary (commutative) configuration space $\{x\}=\{q,p\}$.
The fields $F(q,p)$ and  $\theta(q,p)$   
are then united into one field ${\cal F}(x)$, living in the enlarged space $x$, 
on which the Maxwell equations are imposed. 
(One may deform those equations while extending them to the $p$ part of the space, 
but no compelling reason requires that.)
$\theta$ is then interpreted as a part of an electromagnetic background living in a higher dimensional space.
The relation with noncommutative dynamics is established in the following way.
Consider a particle with mass $M_0$ and position $x(t)$, interacting with the ${\cal F}$ field.
The Lagrangian is
\be
L'=\int dt \left (\frac{M_0}{2}\dot{x}_a^2-{\cal A}_b \dot{x}_b\right )  \label{HD}
\ee
where  ${\cal F}_{ab}(x)=\partial_a {\cal A}_b-\partial_b {\cal A}_a$. $x_a,x_b$ denote both $q_i$ and $p_j$.
To separate between them, use also the notation 
$q_i\equiv x_i$, $q_j\equiv x_j$, $p_i\equiv x_I$, $p_j\equiv x_J$, etc. 
Consider now a background field ${\cal F}$ of the form  
\be
{\cal F}_{ij}=F_{ij}^{-1}(x) \qquad  {\cal F}_{IJ}=\theta_{ij}^{-1}(x) 
\qquad{\cal F}_{iI}=1\qquad{\cal F}_{iJ}=0.
\ee
By a simple extension of a classic argument \cite{peierls} it can be shown that in the limit
\be
M_0\rightarrow 0
\ee
the until now ordinary point particle will transform into one evolving in a phase-space  $\{q_i,p_j\}$
having the 'noncommutative' symplectic structure (\ref{PB_FT}), and driven by a quadratic Hamiltonian
[unless additional terms appear in (\ref{HD})].
According to (\ref{maestru}-\ref{variations}),
a second dimensional reduction may take place, if $F_{ij}\theta_{jk}=-\delta_{ik}$, 

In conclusion, one can interpret the noncommutativity form $\theta$ as a part of a higher
dimensional electromagnetic field ${\cal F}$,
with the noncommutative dynamics (\ref{PB_FT}) of test particles arising upon taking the limit 
$\frac{M_0}{\cal{F}}\rightarrow 0$.


\bigskip

{\bf Acknowledgements}


I am indebted to Drs. K.N. Anagnostopoulos, G.G. Athanasiu, P. Di\c{t}\v{a}, N. Grama and T.N. Tomaras 
for useful discussions.
This work was partially supported through a European Community individual Marie Curie fellowship
under contract HPMF-CT-2000-1060,
and benefitted from the subsequent hospitality of Prof. Tomaras at Crete University.


\end{document}